\documentstyle[12pt]{article}

\newcommand{\ba}{\begin{eqnarray}}
\newcommand{\ea}{\end{eqnarray}}
\input{psfig}
\begin{document}
\pagestyle{plain}
\title{Electromagnetic couplings
in a collective model of the nucleon}
\author{R.~Bijker\\
Instituto de Ciencias Nucleares, U.N.A.M.,\\
Apartado Postal 70-543, 04510 M\'exico D.F., M\'exico
\and
A.~Leviatan\\
Racah Institute of Physics, The Hebrew University,\\
Jerusalem 91904, Israel}
\date{}
\maketitle

\begin{center}
\end{center}

\begin{abstract}
We study the electromagnetic properties of the nucleon and its
excitations in a collective model. In the ensuing algebraic treatment
all results for helicity amplitudes and form factors can be
derived in closed form in the limit of a large model space.
We discuss nucleon form factors and transverse electromagnetic
couplings in photo- and electroproduction, including transition form
factors that can be measured at new electron facilities.
\end{abstract}

\clearpage

\section{Introduction}

The availability and construction of facilities with intense polarized
photons and CW electron accelerators in the GeV and multi-GeV region
with high luminosity beams and large acceptance detectors signals a new era
of hadron spectroscopy exploiting the electromagnetic probe.
Ongoing and planned experiments for exclusive processes and polarization
measurements in light quark baryons are in need of accurate
theoretical interpretation and guidance. This, combined with the lack of
exact solutions of QCD in the nonperturbative regime and the present
inability of lattice calculations to address the entire excitation spectrum,
motivate and necessitate the development of QCD-inspired models for
baryon spectroscopy.

Extensive calculations of baryon observables
were carried out within the quark potential model in its
nonrelativistic \cite{NRQM} or relativized \cite{RQM}
form, which set the standard in the field. Such models emphasize the
single-particle aspects of quark dynamics for which only a few low-lying
configurations in the confining potential contribute significantly.
On the other hand, flux-tube string quark models \cite{flux}, as well
as some regularities in the observed spectra ({\it e.g.} linear Regge
trajectories \cite{regge}, parity doubling \cite{paritydoub}) hint that
an alternative, more collective
type of dynamics may play a role in the structure of baryons.
Algebraic methods are particularly suitable for analyzing collective
forms of dynamics, as demonstrated in nuclear and molecular physics
\cite{IAC}. In this contribution we use such methods to study the
electromagnetic couplings of nonstrange baryons.
Form factors and helicity amplitudes are far more sensitive to details in
wave functions than the mass spectrum, and hence they provide
a good tool to test (and possibly distinguish) different models of
baryon structure.

\section{A collective model of baryons}

In a flux-tube picture low-lying baryon resonances correspond
approximately to three quarks moving in an adiabatic potential
generated by a Y-shaped junction of three flux-tubes \cite{flux}.
With that in mind, we have recently suggested an algebraic
model of baryons \cite{BIL}, in which the nucleon has
the string configuration of Figure~\ref{geometry}.
The string is idealized as a thin string with a distribution of
mass, charge and magnetization $g(\beta)$ where $\beta$ is the
coordinate along the string.
The relevant degrees of freedom of the string configuration of
Figure~\ref{geometry} are the two relative Jacobi coordinates
\ba
\vec{\rho} \;=\; (\vec{r}_1 - \vec{r}_2)/\sqrt{2} ~, \hspace{1cm}
\vec{\lambda} \;=\; (\vec{r}_1 + \vec{r}_2 - 2\vec{r}_3)/\sqrt{6} ~,
\label{jacobi}
\ea
where $\vec{r}_1$, $\vec{r}_2$ and $\vec{r}_3$ denote the end points of
the string configuration. In an algebraic treatment we use instead
second quantization in which the two relative Jacobi coordinates and
their conjugate momenta are replaced by
two vector boson creation ($b^{\dagger}_{\rho}, b^{\dagger}_{\lambda}$)
and annihilation operators ($b_{\rho}, b_{\lambda}$).
The 36 bilinear products of creation and annihilation operators
generate the Lie algebra of $U(6)$, the symmetry group of the
harmonic oscillator quark model \cite{bowlerhey}.
In this case, the model space consists of a
given harmonic oscillator shell labeled by $n=n_{\rho}+n_{\lambda}$.
In order to allow for coupling between various oscillator shells
(needed to achieve collectivity), we
enlarge the model space by adding a scalar boson ($s^{\dagger}, s$),
under the restriction that the total number of bosons
$N=n_{\rho}+n_{\lambda}+n_s$ (vector plus scalar) is conserved.
The 49 bilinear products of the seven (six vector and one scalar)
creation and annihilation operators generate the Lie algebra of
$U(7)$. In this case all states of the model space belong to
the symmetric representation $[N]$ of $U(7)$, which contains the
harmonic oscillator shells with $n=n_{\rho}+n_{\lambda}=0,1,\ldots,N$.
The $s$-boson does not introduce a new degree of freedom because it
satisfies the
constraint $\hat n_{s} = \hat N-\hat n_{\rho}-\hat n_{\lambda}$.
All operators of interest, such as mass operators and transition
operators, are expressed in terms of the generators of $U(7)$.
The full algebraic structure for baryons is obtained by combining the
geometric part with the usual spin-flavor and color parts into the
spectrum generating algebra of
${\cal G} = U(7) \otimes SU_{sf}(6) \otimes SU_c(3)$.

For nonstrange baryons the $S_3$ permutation symmetry between
the identical constituent parts can be incorporated exactly in $U(7)$
\cite{BIL}. The resulting (one- and two-body) $S_3$-invariant mass
operator describe the dynamics of the configuration
in Figure~\ref{geometry}, in which the length of the strings and the
relative angles between them are equal.
In this case the planar shape is an oblate top with $D_{3h}$ point group
symmetry. The classification of states under $D_{3h}$ is equivalent to
the classification under permutations and parity \cite{bunker}.
The eigenstates of the oblate top are characterized by
$(v_1,v_2);K,L_t^{P}$, where $(v_1,v_2)$ denote the
vibrations (stretching, bending); $K$ is the projection of the rotational
angular momentum $L$ on the body-fixed symmetry axis, $P$ the parity
and $t$ the transformation property of the states under $D_3$
(a subgroup of $D_{3h}$), or equivalently the symmetry type under $S_3$.
The antisymmetry of the total baryon wave function implies that the
permutation symmetry of the geometric part is
the same as that of the spin flavor part.
Therefore, there are three equivalent ways
to label the permutation symmetry of the baryon wave function, either
by the dimension of the $SU_{sf}(6)$ representation or by the
representation of the permutation group $S_3$ or by that of the
point group $D_3$:
$56 \leftrightarrow S \leftrightarrow A_1$,
$20 \leftrightarrow A \leftrightarrow A_2$,
$70 \leftrightarrow M \leftrightarrow E$.

In the present approach nonstrange resonances are
identified with rotations and vibrations of a string with the symmetry
of an oblate top. The resulting mass spectrum exhibits rotational
states ($K,L^{P}_{t}$) arranged in bands built on top of each vibration
$(v_1,v_2)$. The corresponding wave functions are spread over many
oscillator shells and hence are truly collective. Although the underlying
dynamics is quite different, the fit for the mass spectrum is
comparable \cite{BIL} to that obtained in quark potential models.

\section{Electromagnetic couplings}

The distinguishing feature between different models of hadrons is
their form factors. This holds especially in the case of baryons whose
size ($\sim 1$ fm) is comparable to the scale of their excitation
energies ($\sim 300$ MeV). For electromagnetic couplings the calculation
of elastic and transition form factors proceeds through a nonrelativistic
reduction of the coupling of point-like constituents to the photon field.
The resulting transition operator contains a nonrelativistic part,
a spin-orbit part, a nonadditive part and higher order corrections \cite{CL}.
In this contribution we limit ourselves to the nonrelativistic part,
which for transverse couplings consists of a magnetic and an electric
contribution
\ba
{\cal H} &=& 6 \sqrt{\pi/k_0} \, \mu e_3 \, \left[ k s_{3,+} \hat U
- \hat T_{+}/g \right] ~.
\label{ht}
\ea
The $\hat z$-axis is taken along the direction of the photon momentum
$\vec{k} = k \hat z$, $k_0$ is the photon energy, and
$e_j$, $\vec{s_j}$ and $\mu=\mu_j=ge/2m_q$ denote the charge, spin
and magnetic moment of the $j$-th constituent.

Helicity ampitudes and form factors of interest in photo- and
electroproduction are proportional to the matrix elements of ${\cal H}$
between initial and final states.
To evaluate these matrix elements in the algebraic model,
the operators $\hat U$ and $\hat T_+$ are expressed in terms of
generators of the $U(7)$ algebra \cite{BIL}
\ba
\hat U &=& \mbox{e}^{ -i k \beta \hat D_{\lambda,z}/X_D } ~,
\nonumber\\
\hat T_+ &=&
\frac{i m_q k_0 \beta}{2 X_D} \left( \hat D_{\lambda,+} \,
\mbox{e}^{ -i k \beta \hat D_{\lambda,z}/X_D } +
\mbox{e}^{ -i k \beta \hat D_{\lambda,z}/X_D } \, \hat D_{\lambda,+}
\right) ~. \label{emop}
\ea
The dipole operator
$\hat D_{\lambda,m} = (b^{\dagger}_{\lambda} \times s -
s^{\dagger} \times \tilde b_{\lambda})^{(1)}_m$, is the
generator of $U(7)$ with the same transformation
properties as the Jacobi coordinate $\lambda_m$.
The coefficient $X_D$ is a normalization factor given by the
reduced matrix element of the dipole operator
$X_D = |\langle 1^-_E || \hat D_{\lambda} || 0^+_{A_1} \rangle|$
and $\beta$ represents a scale of the coordinate.

The matrix elements of the operators in Eq.~(\ref{emop}) can be
obtained by noting that the operator appearing in the exponent is a
generator of $U(7)$. Therefore, the matrix elements of $\hat U$
and $\hat T_+$ can be expressed in terms of group elements of $U(7)$,
which can be evaluated exactly without having to make any further
approximations. There exist limiting situations in which these group
elements can be derived in closed form. In the limit of a large model
space ($N \rightarrow \infty$) we recover for the harmonic oscillator
the familiar expressions in terms of exponentials, and for the oblate
top we find expressions in terms of spherical Bessel functions \cite{BIL}.
In keeping with a collective description of baryons, we assume that the
charge and magnetic moment are not concentrated at the end points of the
string configuration of Figure~\ref{geometry}, but instead are
distributed along the strings. Hereto we fold the matrix
elements of the operators in Eq.~(\ref{emop}) with a probability
distribution for the charge and magnetization of the form
$g(\beta) = (\beta^2/2a^3) \, \mbox{exp}(-\beta/a)$
to obtain the distributed-string or collective form factors
\ba
F(k) &=&  \int \mbox{d} \beta \, g(\beta) \,
\langle \psi_f | \hat U | \psi_i \rangle ~,
\nonumber\\
G_+(k) &=&  \int \mbox{d} \beta \, g(\beta) \,
\langle \psi_f | \hat T_+ | \psi_i \rangle  ~.
\label{radint}
\ea
Here $a$ is a scale parameter.

In Table~\ref{collff} we show the $N \rightarrow \infty$
results for some resonances which in the present approach are
interpreted as rotational excitations of an oblate top.
They are all characterized by $(v_1,v_2)=(0,0)$.
These closed expressions allow one to study the dependence
on momentum transfer. For small values of $k$ (long wavelength limit)
the transition form factors $F(k)$ show the threshold behavior
$\sim k^L$. For large values of $k$ all form factors drop as powers of
$k$. This property is well-known experimentally and is in contrast with
harmonic oscillator quark models in which all form factors fall off
exponentially (see Figure~\ref{protonff}). The elastic form factor
$F(k)$ drops as $k^{-4}$, whereas the transition form factors for all
rotational excitations drop as $k^{-3}$.
The form factors $G_{+}(k)$ drop as the derivatives of $F(k)$.

The analysis of electromagnetic form factors presented in this
contribution is based on the collective form factors of
Table~\ref{collff} which are obtained in the limit of
a large model space ($N \rightarrow \infty$).
The choice for the probability distribution $g(\beta)$ is such that
we recover the familiar dipole form for the elastic form factor $F(k)$.
The value of the scale parameter $a$ in $g(\beta)$ is
determined by the proton charge radius
$a=\langle r^2 \rangle^{1/2}_p=0.249$ fm.

\subsection{Transition form factors}

In photo- and electroproduction of baryon resonances the transverse
couplings are expressed in terms of helicity amplitudes,
$A_{\nu}$ ($\nu=1/2,3/2)$. Their measurement is an essential
part of the research program at current and new electron facilities.
Helicity amplitudes correspond to specific matrix elements of
${\cal H}$,
\ba
A_{\nu} &=& 6 \sqrt{\pi/k_0} \, \mu \, \Bigl[ \beta_{\nu} \, k F(k)
+ \alpha_{\nu} \, G_{+}(k)/g \Bigr] ~, \label{helampt}
\ea
and are all expressed in terms of the radial integrals
of Eq.~(\ref{radint}) and Table~\ref{collff}. The spin-flavor
dependence of the helicity amplitudes is contained in the
$\alpha_{\nu}$ and $\beta_{\nu}$ factors $\cite{BIL}$ and are
common to all models sharing the same spin-flavor structure.

In Figures~\ref{D13} and \ref{F15} we show the transition form factors
for proton couplings to the N$(1520)D_{13}$ and the N$(1680)F_{15}$
resonances. All calculations are done in the Breit frame and include
the sign of the subsequent strong decay of the resonance
$N^{\ast} \rightarrow N + \pi$, which is calculated by assuming
that the meson is emitted from a single constituent.
The collective form factors (solid lines) give a fair
agreement with the data, whereas the harmonic oscillator form factors
(dotted lines) fall off too quickly.
For both resonances, the rapid change in the corresponding asymmetry
parameter $(A_{1/2}^2-A_{3/2}^2)/(A_{1/2}^2+A_{3/2}^2)$
from $-1$ to $+1$ is reproduced. Since the behavior of the asymmetry
parameter is determined by the spin-flavor part and does not depend
on the spatial part, the results obtained with the collective and
the harmonic oscillator form factors are identical.

\subsection{Nucleon form factors}

The elastic electromagnetic form factors of the nucleon can be
evaluated in a similar way. For the electric form factor of the
proton we find (by construction) a dipole form
\ba
G_E^p &=& \frac{1}{(1+k^2a^2)^2} ~. \label{gep}
\ea
In Figure~\ref{protonff} we show the dependence of $G_E^p$
on $Q^2$ ($=k^2$ for elastic transitions).
On the other hand, the neutron electric form factor vanishes
identically in this scheme, $G_E^n=0$. This is a consequence of
the spin-flavor content of the nucleon wave function.

A non-vanishing neutron electric form factor can be obtained by breaking
the $SU_{sf}(6)$ symmetry {\it e.g.} via the hyperfine interaction
\cite{IKS}, or by breaking the $D_3$ spatial symmetry {\it e.g.}
by allowing for a quark-diquark structure \cite{Tzeng} and
flavor-dependent mass terms.
If we assume a flavor dependent probability distribution,
$g_u(\beta) = (\beta^2/2a_u^3) \, \mbox{exp}(-\beta/a_u)$ and
$g_d(\beta) = (\beta^2/2a_d^3) \, \mbox{exp}(-\beta/a_d)$, we obtain
\ba
G_E^p &=& \frac{4}{3(1+k^2a_u^2)^2} - \frac{1}{3(1+k^2a_d^2)^2} ~,
\nonumber\\
G_E^n &=& \frac{2}{3(1+k^2a_u^2)^2} - \frac{2}{3(1+k^2a_d^2)^2} ~.
\label{elff}
\ea
The two scale parameter $a_u=0.258$ fm and $a_d=0.285$ fm
are determined by fitting simultaneously the proton and neutron
charge radii. Figure~\ref{neutronff} shows a comparison with
the neutron electric form factor. The proton electric form factor is
hardly changed when Eq.~(\ref{gep}) is replaced by Eq.~(\ref{elff}).
We note that a relatively small breaking of the flavor symmetry
is sufficient to give a good fit of both the
neutron and the proton electric form factors.
Similarly, for the magnetic form factors we find
\ba
G_M^p &=& \frac{8\mu}{9(1+k^2a_u^2)^2}
+ \frac{\mu}{9(1+k^2a_d^2)^2} ~,
\nonumber\\
G_M^n &=& -\frac{4\mu}{9(1+k^2a_d^2)^2}
- \frac{2\mu}{9(1+k^2a_u^2)^2} ~.
\label{magff}
\ea
In the photoproduction limit the ratio of neutron and
proton magnetic form factors is that of the corresponding magnetic
moments $\mu_n/\mu_p=-0.685$ which agrees very well with the
value $-2/3$ calculated from Eq.~(\ref{magff}).
On the basis of perturbative QCD one
expects that for large values of $Q^2$ the above ratio approaches
$-1/2+{\cal O}(\mbox{ln}Q^2)$ \cite{WPR}.
With the same values of the scale parameters as for the electric form
factors we calculate this ratio to be $-0.54$,
in agreement with the p-QCD value.
Without the flavor dependence in the probability distribution
($a_u=a_d$) the ratio is $-2/3$, independent of $Q^2$.

\section{Summary, conclusions and outlook}

We have presented a study of electromagnetic couplings
of baryon resonances in a collective model of the nucleon. All
helicity amplitudes and form factors of interest can be expressed
in closed form in the limit of a large model space
($N \rightarrow \infty$).
We found a consistent description of photo- and electroproduction
of the experimentally well-known N$(1520)D_{13}$ and N$(1680)F_{15}$
resonances. An application to the nucleon elastic form factors showed that
by introducing a flavor dependent probability distribution we could
simultaneously describe the proton and neutron electric form
factors as well as the change in the ratio of their
magnetic form factors.

Work on the calculation of photo- and electroproduction
of all low lying resonances including the effects of higher
order terms in the transition operator and the extension to
longitudinal couplings is currently in progress.

\section*{Acknowledgements}

This work was supported in part by DGAPA-UNAM under project IN105194
and by the Basic Research Foundation of the
Israel Academy of Sciences and Humanities.

\clearpage

\begin{table}
\centering
\caption[]{\small
Collective form factors of Eq.~(\ref{radint}) for large $N$.
$H(x)=\arctan x - x/(1+x^2)$. The final states are
labeled by $[f,L^P]_{(v_1,v_2);K}$, where $f$ denotes the
dimension of the $SU_{sf}(6)$ representation.
The initial state is $[56,0^+]_{(0,0);0}$.
\normalsize}
\label{collff}
\vspace{15pt}
\begin{tabular}{ccc}
\hline
& & \\
Final state & $F(k)$ & $G_+(k)/m_q k_0 a$ \\
& & \\
\hline
& & \\
$[56,0^+]_{(0,0);0}$
& $\frac{1}{(1+k^2a^2)^2}$ & 0  \\
& & \\
$[70,1^-]_{(0,0);1}$ & $-i \, \sqrt{3} \, \frac{ka}{(1+k^2a^2)^2}$
& $\sqrt{2}F(k)/ka $ \\
& & \\
$[56,2^+]_{(0,0);0}$
& $ \frac{1}{2} \sqrt{5}\left[ \frac{-1}{(1+k^2a^2)^2}
+ \frac{3}{2k^3a^3} H(ka) \right]$
& $\sqrt{6}F(k)/ka $ \\
& & \\
$[70,2^+]_{(0,0);2}$
& $-\frac{1}{2} \sqrt{15}\left[ \frac{-1}{(1+k^2a^2)^2}
+ \frac{3}{2k^3a^3} H(ka) \right]$
& $\sqrt{6}F(k)/ka $ \\
& & \\
\hline
\end{tabular}
\end{table}

\clearpage

\begin{figure}
\caption[]{Collective model of baryons and its idealized string
configuration (the charge distribution of the proton is shown as
an example).}
\label{geometry}
\end{figure}

\begin{figure}
\caption[]{Helicity amplitudes $A^p_{\nu}$ in $10^{-3}$ (GeV)$^{-1/2}$
for the N$(1520)D_{13}$ resonance. The curves correspond to the
collective form factor (solid lines) and the harmonic oscillator
form factor (dotted lines). The experimental data are taken from
the compilation in \cite{Burkert}.}
\label{D13}
\end{figure}

\begin{figure}
\caption[]{Same as Figure~\ref{D13},
but for the N$(1680)F_{15}$ resonance.}
\label{F15}
\end{figure}

\begin{figure}
\caption[]{Comparison between the experimental proton electric
form factor $G_E^p$, the collective form factor (solid lines)
and the harmonic oscillator form factor (dotted lines).
The experimental data, taken from a compilation in
\cite{gep}, and the calculations are divided by the dipole form
factor, $F_D=1/(1+Q^2/0.71)^2$.
The curves are obtained a) from Eq.~(\ref{gep}) and the
harmonic oscillator form factor $\mbox{exp}(-k^2 \beta^2/6)$
with $\beta=a \sqrt{12}$, and b) from Eq.~(\ref{elff}) and the
corresponding expression for the harmonic oscillator form factor
with $\beta_u=a_u \sqrt{12}$ and $\beta_d=a_d \sqrt{12}$.}
\label{protonff}
\end{figure}

\begin{figure}
\caption[]{Comparison between the experimental neutron electric
form factor $G_E^n$, the collective form factor (solid line) and
the harmonic oscillator form factor (dotted line).
The experimental data are taken from a compilation in
\cite{gen}.}
\label{neutronff}
\end{figure}

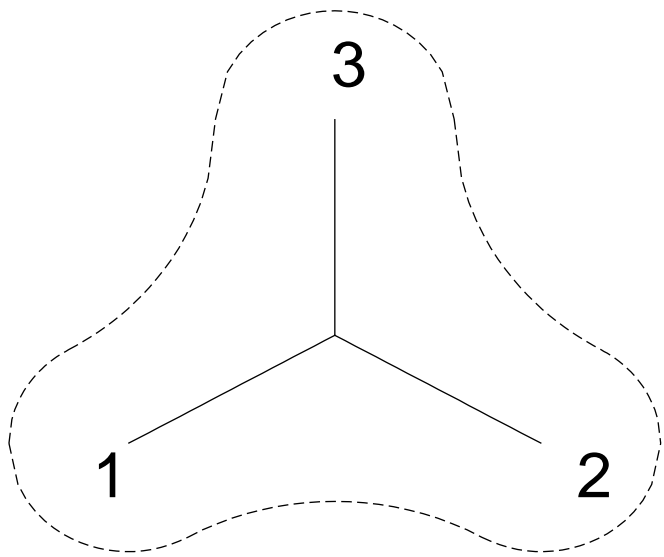
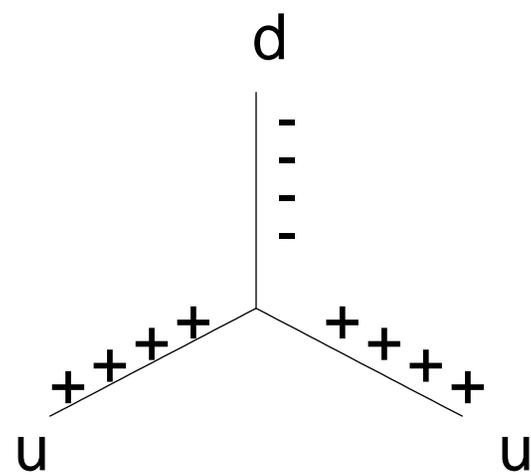
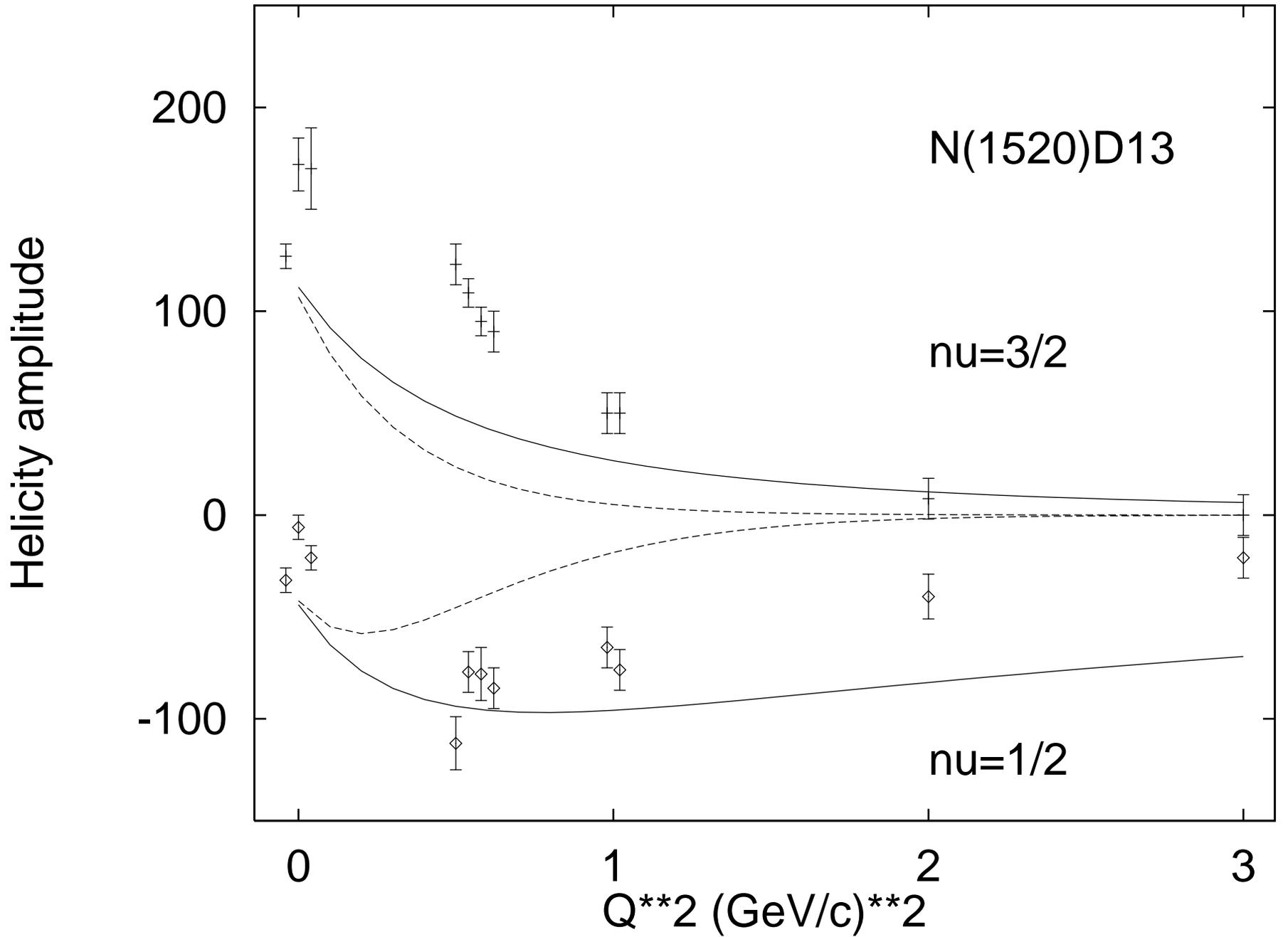
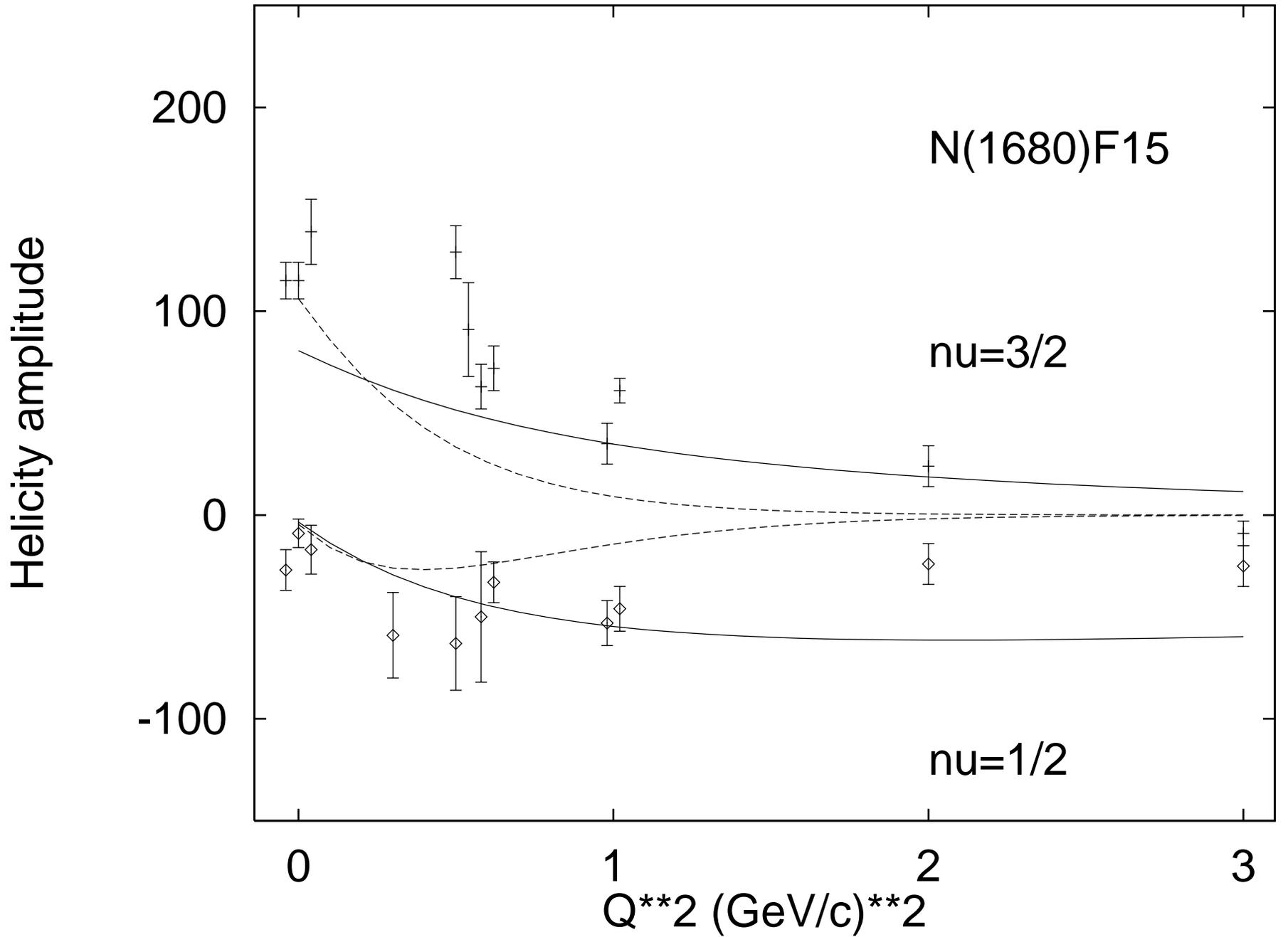
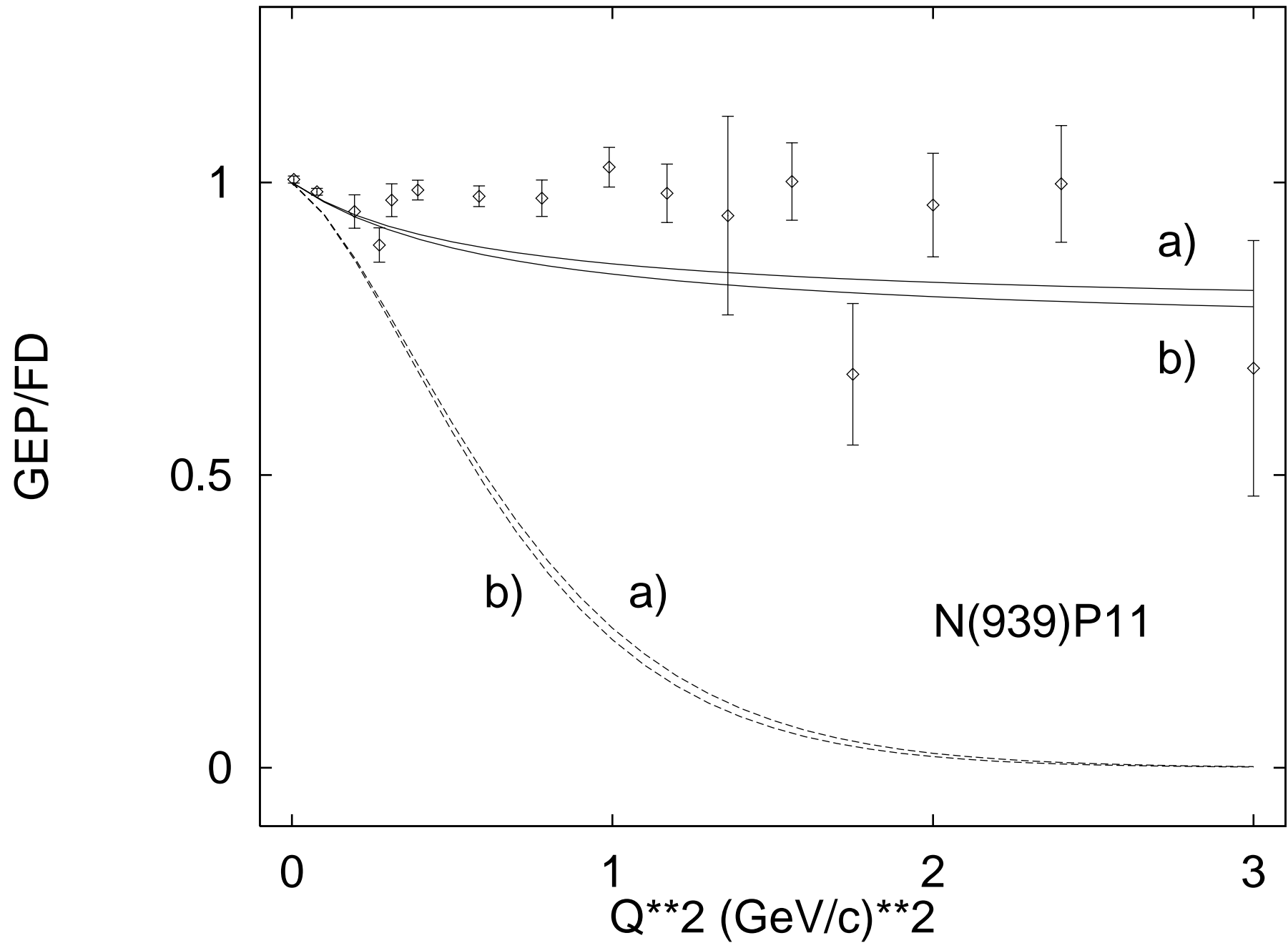
\end{document}